\documentclass[aps,pra,reprint,amsmath,amssymb,longbibliography,nofootinbib]{revtex4-2}

\usepackage{soul}
\usepackage[utf8]{inputenc}
\usepackage{bm,amsfonts,amsmath,amssymb,color}
\usepackage{dcolumn,xfrac}
\usepackage{graphicx}
\def\la{\;
\raise0.3ex\hbox{$<$\kern-0.75em\raise-1.1ex\hbox{$\sim$}}\; }
\def\ga{\;
\raise0.3ex\hbox{$>$\kern-0.75em\raise-1.1ex\hbox{$\sim$}}\; }

\newcommand{\dmm}{$\Delta\mu/\mu$}

\newcommand{\kms}{km~s$^{-1}$}

\newcommand{\etal}{{et al.}}

\begin{document}
\title{Spectroscopic Shifts in Deuterated Methanol Induced by Variation of $m_{\rm e}/m_{\rm p}$}


\author{J. S. Vorotyntseva$^{1,2}$}
\author{S. A. Levshakov$^{1}$} 
\author{M. G. Kozlov$^{2,3}$}

\affiliation{
$^{1}$ Ioffe Institute, Polytekhnicheskaya Str. 26, 194021 St.~Petersburg, Russia\\
$^{2}$ St.~Petersburg Electrotechnical University ``LETI'', Prof. Popov Str. 5, 197376 St.~Petersburg, Russia\\
$^{3}$ Petersburg Nuclear Physics Institute of NRC ``Kurchatov Institute'', Gatchina, Leningrad District, 188300, Russia
}

\begin{abstract}
Numerical calculations of the sensitivity coefficients, $Q_\mu$, of microwave molecular transitions in the ground torsion-rotation state of deuterated methanol (CH$_3$OD, CD$_3$OH, and CD$_3$OD) to small variations in the fundamental physical constant $\mu = m_{\rm e}/m_{\rm p}$~-- the electron-to-proton mass ratio~-- are reported. Theoretical motivation for changes in $\mu$ comes from a variety of models beyond the Standard Model of particle physics which are
invoked to explain the nature of dark matter and dark energy that dominate the Universe. The calculated values of $Q_\mu$ range from $-300$ to +73 and, thus, make deuterated methanol promising for searches for small space-time changes in $\mu$. 
It is also shown that among the calculated sensitivity coefficients $Q_\mu$ using different Hamiltonians in the present and previous works, there are several pronounced outliers of unclear nature.
\end{abstract}

\date{\today}

\maketitle


\section{Introduction}

\label{Sec1}

Since the pioneering work of [1]
the $\mu$-test has been widely
used in experimental studies of theoretical models 
beyond the Standard Model of particle physics which
predict small variations of fundamental physical constants and, in particular,
the electron-to-proton mass ratio, $\mu = m_{\rm e}/m_{\rm p}$,
in space and time.
Theories that connect varying constants to dark matter consider the
dependence of $\mu$ on the local density of dark matter which differs
across the Milky Way ([2]). 
The coupling of physical constants to either dark matter, or to gravitational fields is related
to the chameleon scenario (e.g., [3], [4])
and it can be probed in the Milky Way, or in the nearby galaxies
by astrophysical observations
([5], [6]).

Regular quantitative measurements of fractional changes in $\mu$ between 
its extraterrestrial ($\mu_{\scriptscriptstyle\rm obs}$) 
and terrestrial ($\mu_{\scriptscriptstyle\rm lab}$) values,
${\Delta\mu}/{\mu}= 
(\mu_{\scriptscriptstyle\rm obs}-\mu_{\scriptscriptstyle\rm lab})/\mu_{\scriptscriptstyle\rm lab}$,
became possible after the identification of molecular hydrogen lines in the 
absorption-line spectra of distant quasars ([7])
and calculation of sensitivity coefficients, 
$Q_\mu$, for individual molecular transitions of the Lyman and Werner bands of H$_2$
([8]).

It should be noted that at the first stage these studies were carried out mainly in the optical range on
the H$_2$ lines for which the sensitivity coefficients are not very high, $|Q_\mu| \sim 10^{-2}$,
and the spectral resolution (channel width) does not exceed 1~\kms.
The tightest upper limits on the variability of $\mu$ on the cosmological time scale were obtained in the
framework of the VLT/UVES Large Programme for testing fundamental physics,
\dmm~$< 10^{-5}$ ([9]),
and at the highest redshift $z = 4.22$ towards the quasar
J1443+2724, \dmm~$< 8\times10^{-6}$ ([10])
\footnote{All upper limits on \dmm\ throughout
the paper are given at the confidence level of $1\sigma$.}.

The second stage of development of the $\mu$-tests took place in the microwave radio range after
[11]
showed the enhanced sensitivity of the inversion tunneling transition
at 23~GHz in ammonia molecule NH$_3$, $Q_\mu = +4.46$.
Comparing the positions of the ammonia inversion transition with pure rotation transitions in
CO, HCO$^+$, and HCN, for which $Q_\mu = 1$, they obtained a constraint on the changes in $\mu$
at redshift $z = 0.6847$ towards the quasar B0218+357, \dmm~$< 2\times10^{-6}$.

The accuracy of \dmm\ estimates depends linearly on the accuracy of position measurements of spectral lines
and is inversely proportional to the difference in the sensitivity coefficients of a pair of lines
used in the analysis ([12]):
\begin{equation}
\frac{\Delta\mu}{\mu}= \frac{V_j-V_i}{c(Q_{\mu,i}-Q_{\mu,j})},
\label{Eq1-1}
\end{equation}
where $V_j$ and $V_i$ are the local standard of rest radial velocities, $V_{\rm LSR}$,
of a pair of molecular transitions having sensitivity coefficients $Q_{\mu,j}$ and
$Q_{\mu,i}$, and $c$ is the speed of light.

The next step in improving \dmm\ estimates was the search for molecules with large differences
in the sensitivity coefficients of one molecule to avoid spectroscopic shifts 
(the so-called Doppler noise)
caused by segregation 
effects in the spatial distribution of molecules within molecular clouds when 
\dmm\ is calculated by
comparing the positions of the lines of different molecules.
The methanol molecule CH$_3$OH turned out to be most suitable for this purpose.
The parent methanol and its isotopologues are among the most abundant molecules in interstellar space.
The sensitivity coefficients for CH$_3$OH lie in the range $-53 \la Q_\mu \la +43$ 
([12], [13])
which made it possible to obtain restrictions on
cosmological variations of $\mu$ at the level of $5\times10^{-8}$ at $z = 0.88582$ towards the
quasar PKS1830-211 ([14]).

The same order of magnitude constraints on the spatial $\mu$-variation in the Milky Way disk 
over the range of the galactocentric distances $4 \la R \la 12$ kpc
are based on the methanol thermal emission lines ([15])
and methanol masers observations ([16], [17]):
\dmm~$< (2-3)\times10^{-8}$. 

It should be emphasized that over the past five decades, improvements in \dmm\ estimates
have always followed improvements in the quality of astronomical equipment and corresponded
to the limiting parameters of observations in optical and radio ranges
(see, e.g., Fig.~1 in [18]).
Such extreme measurements may introduce unaccounted systematic errors into the final results.
Therefore, to obtain robust results, it is necessary to use various instruments and expand the
list of molecular transitions with high sensitivity to changes in $\mu$.
Among a number of complex organic molecules studied (CH$_3$COH, CH$_3$CONH$_2$, 
CH$_3$OCOH, CH$_3$COOH in [13];
CH$_3$NH$_2$ in [19];
and CH$_3$SH in [20]),
suitable transitions were found in methanol isotopologues ([21]):
for 8 lines of deuterated methanol CD$_3$OH
from the frequency interval $1.202 < \nu < 20.089$ GHz, the sensitivity
coefficients\footnote{Note that $Q_\mu=-K_\mu$,
where $K_\mu$ is used by [21]
who defined $\mu$ as the $proton$-$to$-$electron$ mass ratio.}
lie between $-330$ and +88;
2 lines of CD$_3$OD at 2.238~GHz and 2.329~GHz have, respectively, $Q_\mu = -45$ and +80;
2 lines of CH$_3$$^{18}$OH at 2.605~GHz and 11.630~GHz have, respectively, $Q_\mu = -93$ and +34;
and $Q_\mu = +63$ for one line of $^{13}$CH$_3$OH at 1.990~GHz.
 
A list of lines with high sensitivity coefficients of the
$^{13}$C- and $^{18}$O-bearing methanol has recently been expanded
to 27 transitions of $^{13}$CH$_3$OH from the frequency interval 
$1.989 < \nu < 75.415$ GHz with $Q_\mu$ ranging from $-32$ to +78, and
to 30 transitions of CH$_3$$^{18}$OH with $2.604 < \nu < 105.181$ GHz and 
$-109 \leq Q_\mu \leq +33$ ([22]).
We have used these newly calculated $Q_\mu$ values for $^{13}$CH$_3$OH to set constraint
on \dmm~$< 3\times10^{-8}$ from
the thermal emission lines of $^{13}$CH$_3$OH detected 
in the star-forming region NGC~6334I by [23].

Astrophysical observations of deuterated methanol in Galactic molecular clouds
have been performed previously and are still an important goal of current studies.
The CH$_3$OD lines were detected in the high-mass star-forming regions Sgr~B2
([24], [25], [26])
and 
Orion-KL 
([27], [28], [29], [30])
which are both the reference sources to search for
complex organic molecules.
Besides the isotopologue CH$_3$OD was observed in the high-mass star-forming regions NGC~7538-IRS1
([31]), 
NGC~6334I 
([32]),
DR~21 ([33]),
in the intermediate-mass star-forming regions CepE-mm and OMC2-FIR4
([34]),
and in the low-mass star-forming regions IRAS~16293-2422
([35], [36]),
IRAS~4A, IRAS~4B, and IRAS~2
([37]),
and SVS~13A 
([38]),
and HH~212 ([39]).
Triply deuterated methanol, CD$_3$OH, was observed in the low-mass protostar 
IRAS 16293-2242 ([40]).
The search for quadruply deuterated methanol, CD$_3$OD, towards this object has not yet led to
a reliable result but the research is currently ongoing ([41]).

The present work is a continuation of our calculations of sensitivity coefficients
for deuterated methanol CH$_3$OD, CD$_3$OH, and CD$_3$OD in a wider spectral range
as compared to the previous work by [21].

\section{Calculations}
\label{Sec2}

Each of the deuterated methanol molecules
(CH$_3$OD, CD$_3$OH, and CD$_3$OD) has, along with the overall rotation,
internal torsion motions, the energy spectrum of which is similar to that of methanol CH$_3$OH.
The geometrical structure of these molecules is characterized by slight asymmetry which is
the same for CH$_3$OH and CD$_3$OH, but increases for CH$_3$OD and CD$_3$OD resulting from
the off-axis deuterium. 
Small differences in the degree of asymmetry between CH$_3$OH and deuterated methanol molecules
do not, however, affect the results of calculations of the energy spectra of deuterated molecules
using the torsion-rotation Hamiltonian models developed for the parent methanol, as shown 
by [42]
who demonstrated that for a particular molecule CH$_3$OD
a global fit of the observed transitions lies within the experimental uncertainties
(a root-mean-square deviation of 0.098 MHz for $J \leq 10$).
Therefore, when calculating the sensitivity coefficients $Q_\mu$ for lines in CH$_3$OD, CD$_3$OH, and CD$_3$OD,
we used the developed method from our previous works in which the sensitivity coefficients were
calculated for CH$_3$OH ([12])
and for two isotopologues $^{13}$CH$_3$OH and CH$_3$$^{18}$OH
([22]).

\begin{figure}[htb]
\includegraphics[width=0.6\textwidth]{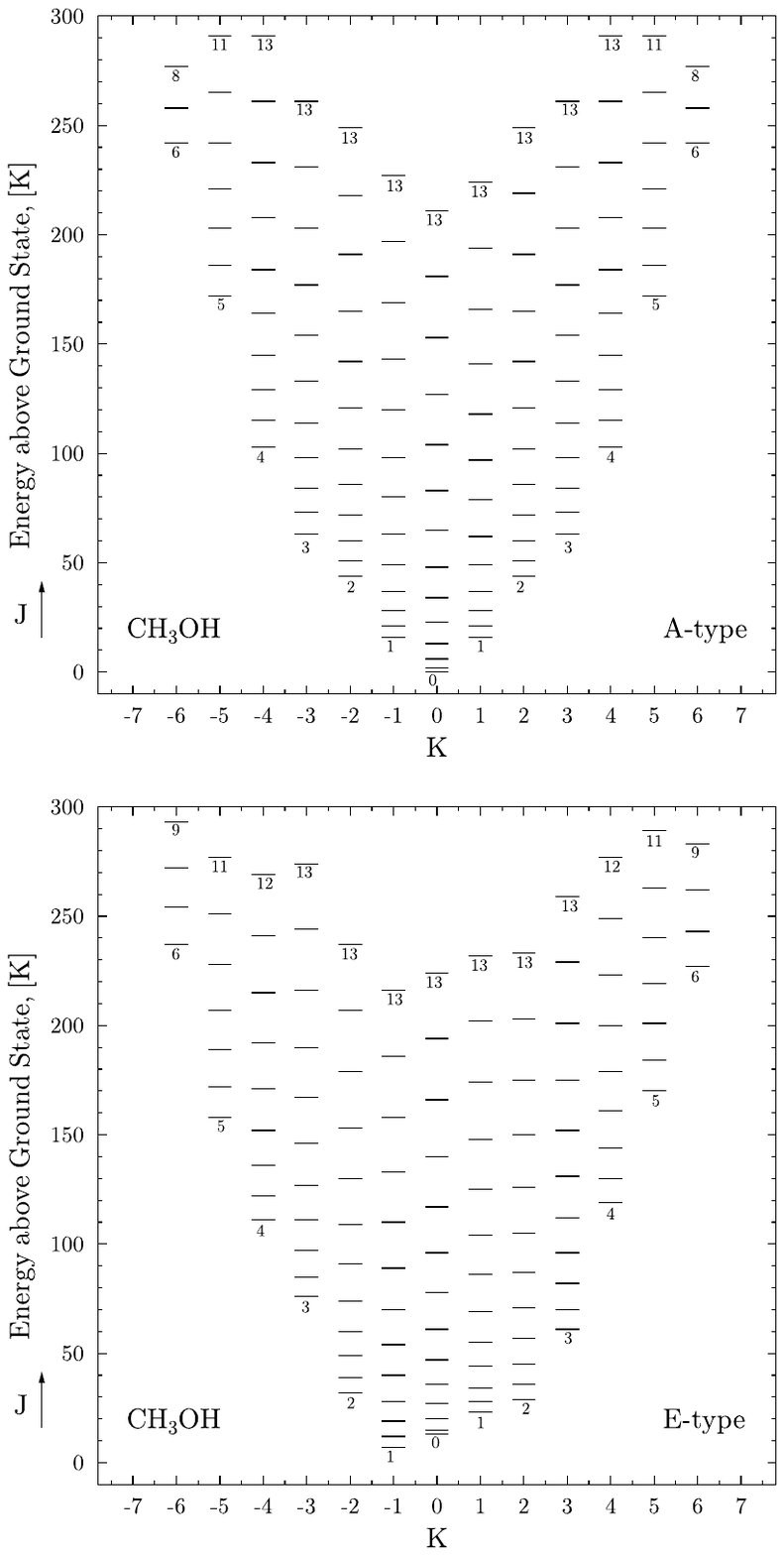}
\caption{
Partial rotational level diagrams of $A$- and $E$-type methanol CH$_3$OH 
in the torsional ground state ($v_t = 0$). 
Numbers below the horizontal bars are the values of the rotational
angular momenta $J$.
}
\label{F1}
\end{figure}

\begin{figure}
\includegraphics[width=0.6\textwidth]{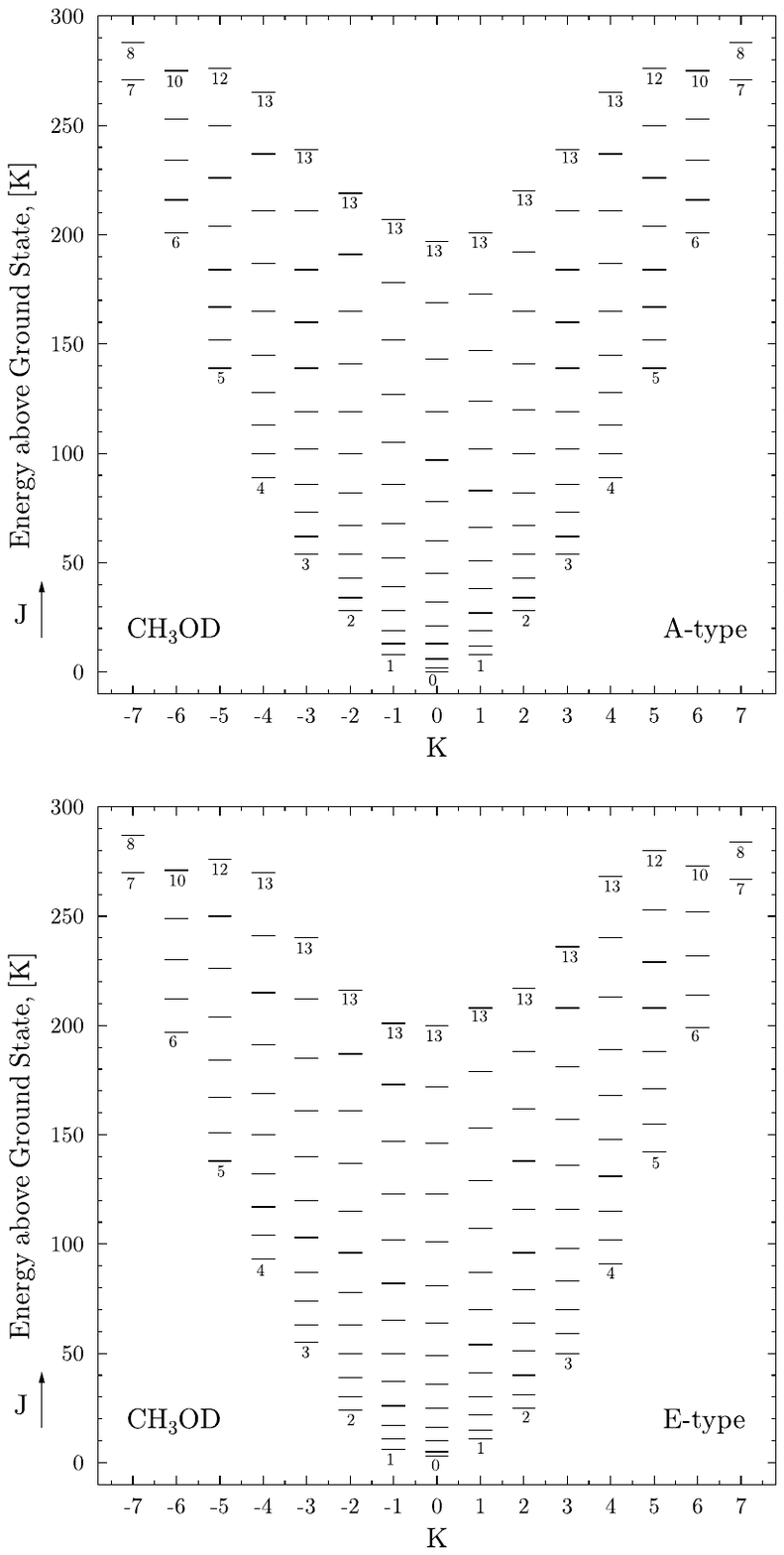}
\caption{
Same as Fig.~\ref{F1} but for the deuterated isotopologue CH$_3$OD.
}
\label{F2}
\end{figure}

\begin{figure}
\includegraphics[width=0.6\textwidth]{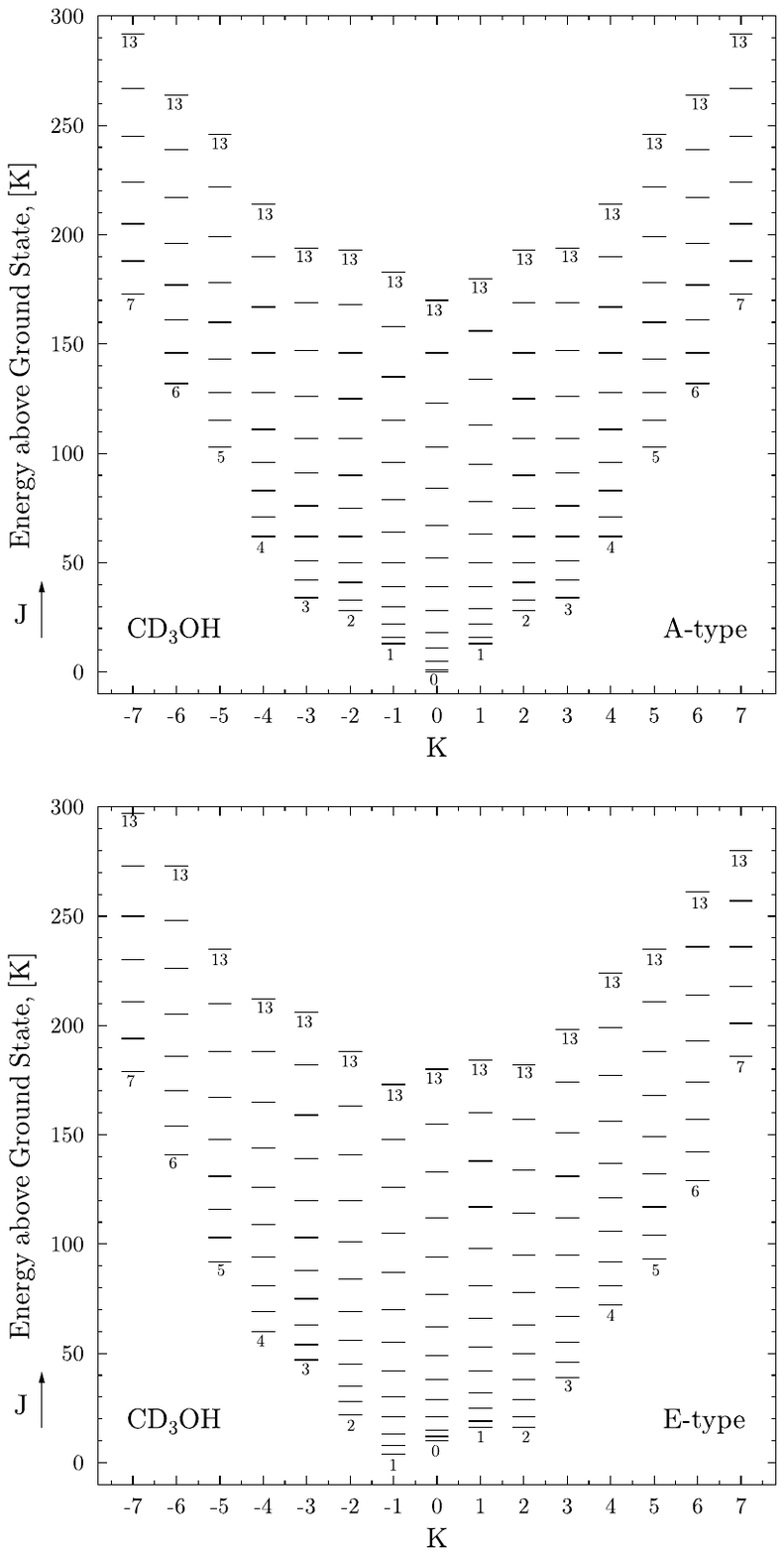}
\caption{
Same as Fig.~\ref{F1} but for the deuterated isotopologue CD$_3$OH.
}
\label{F3}
\end{figure}

\begin{figure}
\includegraphics[width=0.6\textwidth]{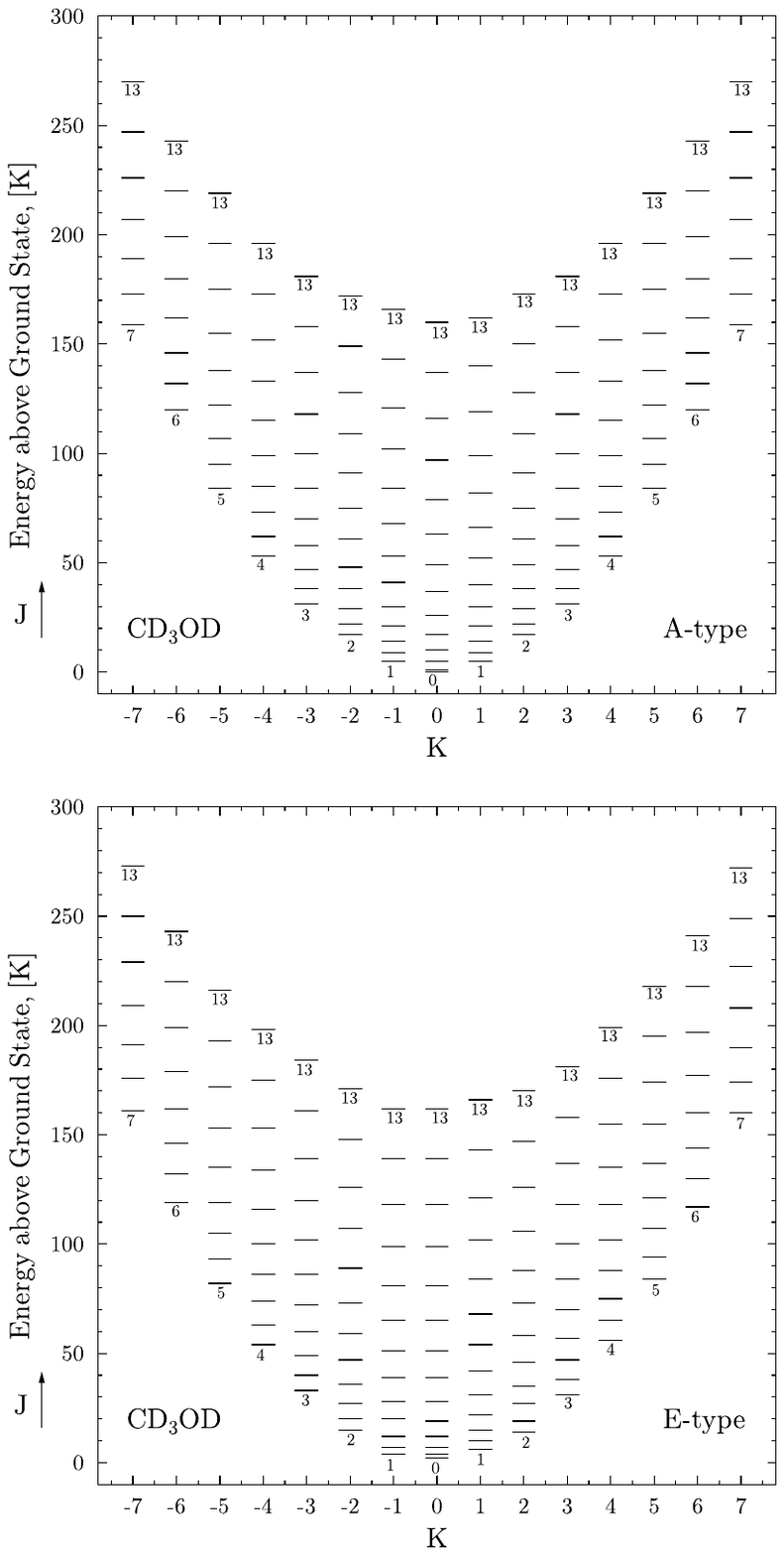}
\caption{
Same as Fig.~\ref{F1} but for the deuterated isotopologue CD$_3$OD.
}
\label{F4}
\end{figure}

\subsection{Sensitivity coefficients and hyperfine corrections}
\label{SSec2-1}

Methanol in nature occurs in two types~-- $A$ and $E$, which differ in the values of the
total nuclear spin of the three hydrogen atoms in the methyl group CH$_3$.
The $A$-methanol has parallel proton spins and the total nuclear spin $I_{123} = 3/2$, whereas 
the $E$-methanol
has one of the protons of the methyl group an antiparallel spin with respect to the
others and $I_{123} = 1/2$. These non-zero total nuclear spins, interacting with the
magnetic fields generated by the overall and torsion rotations of the molecule, lead to
a hyperfine splitting of energy levels.  

As noted by [43],
the hyperfine splitting of the torsion-rotation levels of
molecules with hindered internal rotation such as methanol and its deuterated isotopologues
may affect values of the sensitivity coefficients $Q_\mu$, since hyperfine effects
shift the centre of torsion-rotation lines which was not accounted for 
by effective torsion-rotation Hamiltonians used in previous calculations of $Q_\mu$ 
in [12], [21].
Below we outline conditions when the hyperfine structure of the torsion-rotation levels
can be neglected.

The frequency of a specific hyperfine transition, $\nu$, 
between upper and lower states consists of the rotation, $\nu_{\scriptscriptstyle r}$,
torsion, $\nu_{\scriptscriptstyle t}$, and hyperfine, $\nu_{\scriptscriptstyle h}$, parts
\begin{equation}
\nu = \nu_{\scriptscriptstyle r} + \nu_{\scriptscriptstyle t} + \nu_{\scriptscriptstyle h}\ , 
\label{Eq2-1}
\end{equation}
which are scaled with $\mu$ as $\mu^1$, $\mu^\tau$, and $\mu^2$, respectively,
where $\tau$ is a number attributed to a given transition.

The sensitivity coefficient $Q_\mu$ is determined by two derivatives
\begin{equation}
\frac{d\nu}{\nu} = Q_\mu \frac{d\mu}{\mu}\ .
\label{Eq2-2}
\end{equation}

From (\ref{Eq2-1}) it follows that
\begin{equation}
\frac{d\nu}{d\mu} = \frac{1}{\mu}(\nu_{\scriptscriptstyle r} + \tau \nu_{\scriptscriptstyle t} +
2\nu_{\scriptscriptstyle h})\ ,
\label{Eq2-3}
\end{equation}

\noindent
and then
\begin{equation}
\frac{d\nu}{\nu} = \frac{d\mu}{\mu}\left[
1 + (\tau-1)\frac{\nu_{\scriptscriptstyle t}}{\nu} + \frac{\nu_{\scriptscriptstyle h}}{\nu} \right]\ ,
\label{Eq2-4}
\end{equation}

\noindent
or, using definition of $Q_\mu$, we have
\begin{equation}
Q_{\mu} \equiv Q^{(\scriptscriptstyle r)}_{\mu} + Q^{(\scriptscriptstyle t)}_{\mu} + Q^{(\scriptscriptstyle h)}_{\mu} =
1 + (\tau-1)\frac{\nu_{\scriptscriptstyle t}}{\nu} + \frac{\nu_{\scriptscriptstyle h}}{\nu}\ .
\label{Eq2-5}
\end{equation}

Therefore, for a pure rotation transition, the sensitivity coefficient 
$Q^{(\scriptscriptstyle r)}_{\mu} = 1$.
The energy difference between torsion-rotation levels of methanol in the microwave range
is of the order of $\sim 10$s GHz, whereas the hyperfine splitting occurs at $\sim 10$s kHz,
so that $|\nu_{\scriptscriptstyle h}/\nu| \ll 1$ and, thus, we can neglect the hyperfine
corrections in the effective Hamiltonian.

If, however, the hyperfine splitting is large compared to the transition frequency,
then the 3rd term in (\ref{Eq2-1}) and (\ref{Eq2-5}) must be included in calculations 
of the sensitivity coefficients.
In this case the torsion-rotation transition splits into several lines with different frequencies
and different $Q_\mu$ values.

As for the 2nd term in (\ref{Eq2-1}) and (\ref{Eq2-5}), it shows that 
transitions with enhanced sensitivities occur 
when $\nu_{\scriptscriptstyle t}$ and $\nu_{\scriptscriptstyle r}$ have different signs 
(i.e., energies associated with the hindered internal rotation and the overall rotation 
are canceled) and $|\nu_{\scriptscriptstyle t}/\nu| \gg 1$.
We note in passing that different methanol isotopologues have slightly different rotation $\nu_{\scriptscriptstyle r}$
and torsion $\nu_{\scriptscriptstyle t}$ frequencies, but significantly different $\nu$. 
This results in different $Q_\mu$ values for transitions with identical sets of quantum numbers 
in methanol isotopologues.

The microwave torsion-rotation transition can be defined by the quantum numbers $J$ and $K$
which are, respectively, the total angular momentum and its projection on the axis of the molecule.
If subscripts $u$ and $\ell$ will denote the upper and lower energy levels,
then the microwave torsion-rotation transition in emission can be written as
$J_{{\scriptscriptstyle u}K_{\scriptscriptstyle u}} \to 
J_{{\scriptscriptstyle \ell}K_{\scriptscriptstyle \ell}}$.
Figures~\ref{F1}-\ref{F4} show diagrams
of the methanol and its deuterated isotopologues energy levels
in the torsional ground state ($v_t = 0$) 
which range up to an excitation temperature of 300~K above ground.
The allowed transitions occur in accord with selection rules which have the following forms 
for $A$- and $E$-type methanol (e.g., [44]):

\begin{tabular}{l l c}
\multicolumn{3}{l}{\underline{$A$-{\rm type}}} \\
$\Delta J = 0$ & $\Delta K = 0, \pm1$ & \,\,\, $\pm \leftrightarrow \mp $\\ 
$\Delta J = \pm1$ & $\Delta K = 0, \pm1$ & \,\,\, $\pm \leftrightarrow \pm $ 
\end{tabular}

\begin{tabular}{l l}
\multicolumn{2}{l}{\underline{$E$-{\rm type}}} \\
$\Delta J = 0$ & $\Delta K = \pm1$  \\
$\Delta J = \pm1$ & $\Delta K = 0, \pm1$  
\end{tabular}

\bigskip\noindent
It is interesting to note the inversion of energy levels in some
transitions with identical sets of quantum numbers ($J$, $K$) when
passing from one isotopologue to another which affects the values of $Q_\mu$ considerably.
For instance, the transition
$J_{{\scriptscriptstyle u}K_{\scriptscriptstyle u}} \to 
J_{{\scriptscriptstyle \ell}K_{\scriptscriptstyle \ell}}
= 2_1-3_0 A^+$
in CH$_3$OH has $Q_\mu = +6.3$, but its inverted analog
$J_{{\scriptscriptstyle u}K_{\scriptscriptstyle u}} \to 
J_{{\scriptscriptstyle \ell}K_{\scriptscriptstyle \ell}}
= 3_0-2_1 A^+$ in CH$_3$OD has $Q_\mu = -19.3$, or
$J_{{\scriptscriptstyle u}K_{\scriptscriptstyle u}} \to 
J_{{\scriptscriptstyle \ell}K_{\scriptscriptstyle \ell}}
= 4_{-1}-3_0 E$
in CH$_3$OH has $Q_\mu = -9.6$, but
$J_{{\scriptscriptstyle u}K_{\scriptscriptstyle u}} \to 
J_{{\scriptscriptstyle \ell}K_{\scriptscriptstyle \ell}}
= 3_0-4_{-1} E$ in CD$_3$OH has $Q_\mu = +73$, etc.

\subsection{Calculating procedure}
\label{SSec2-2}

The procedure for calculating sensitivity coefficients of methanol lines from microwave range
was developed in [12].
It is based on an approach by [45]
who suggested
a convenient form of the effective Hamiltonian (hereafter RF Hamiltonian) which
contains six spectroscopic constants having clear dependence on $\mu$.
Comparison with the results of calculations using a more complex form of the Hamiltonian
from [21]
showed that the RF Hamiltonian gives quite
accurate values of the sensitivity coefficients 
([12], [22]).

Seven parameters of the RF Hamiltonian are three rotation parameters $A$, $B$, $C$,
one parameter $D$, describing interaction of the internal rotation with overall rotation,
the kinetic coefficient $F$, and the depth of the three-fold symmetric torsion potential $V_3$,
\begin{equation}
V(\omega)=\frac{V_3}{2}(1 - \cos3\omega)\ ,
\label{Eq2}
\end{equation}
where $0 \le \omega \le 2\pi$ is the torsion angle of the internal rotation of the CH$_3$ group
relative to the OH radical.

The values of all parameters, except $V_3$, were calculated from the moments of inertia
of CH$_3$OD, CD$_3$OH and CD$_3$OD listed in Table~\ref{T1}.
The spectroscopic constants are defined as follows:
\begin{equation}
A = \frac{1}{2}\hbar^2 \left(\frac{I_a+I_b}{I_aI_b-I^2_{ab}}-\frac{I_b}{I^2_b+I^2_{ab}}\right),
\label{Eq3}
\end{equation}

\begin{equation}
B = \frac{1}{2}\hbar^2 \frac{I_b}{I^2_b+I^2_{ab}},
\label{Eq4}
\end{equation}

\begin{equation}
C = \frac{1}{2}\hbar^2 \frac{1}{I_c},
\label{Eq5}
\end{equation}

\begin{equation}
D = \frac{1}{2}\hbar^2 \frac{I_{ab}}{I^2_b+I^2_{ab}},
\label{Eq6}
\end{equation}

\begin{equation}
F = \frac{1}{2}\hbar^2 \frac{I_a I_b - I^2_{ab}}{I_{a2}(I_{a1}I_{b}-I^2_{ab})}.
\label{Eq7}
\end{equation}
Here $\hbar=h/2\pi$, $I_a$, $I_b$, $I_c$ are the moments of inertia, $I_{ab}$
is the product of inertia about the $a$- and $b$-axes in the $a,b,c$-axis system whose $a$-axis is parallel
to the internal rotation axis (assumed to be that of the methyl top), the $c$-axis perpendicular to the COH plane.
$I_{a2}$ is the axial moment of inertia of the methyl group, and $I_{a1}$ is that of the framework 
(the hydroxyl group), the sum of which determines $I_{a}$:
$I_{a} = I_{a1} + I_{a2}$ (for details, see [46]).
The RF Hamiltonian also includes a dimensionless
parameter $\rho$~-- the internal rotation interaction constant,
which is given for asymmetric molecules by
\begin{equation}
\rho =  \frac{I_{a2}\sqrt{I^2_b+I^2_{ab}}}{I_{a}I_{b}-I^2_{ab}}.
\label{Eq8}
\end{equation}

\begin{table}
\caption{Moments of inertia (in units amu$\cdot$\AA$^2$) for the parent methanol molecule
and its deuterated isotopologues from [47].}
\label{T1}
\begin{tabular}{l r@.l r@.l r@.l r@.l}
\hline\\[-5pt]
 & \multicolumn{2}{c}{CH$_3$OD} & \multicolumn{2}{c}{CD$_3$OH}
& \multicolumn{2}{c}{CD$_3$OD} & \multicolumn{2}{c}{CH$_3$OH}   \\
\hline\\[-7pt]
$I_{a1}$ & 1&3946  &0&7514     &1&4082 &0&7493\\
$I_{a2}$ & 3&2249  &6&3872     &6&3939&3&2122\\
$I_{a}^*$ & 4&6195 &7&1386     &7&8021&3&9615\\
$I_{b}$ & 21&5304 &25&4621    &26&7242&20&4771\\
$I_{ab}$ & 0&850 &$-0$&125  &0&8047&$-$0&065\\
$I_{c}$ & 22&9880  &26&2203     &28&1529&21&2614\\
\hline\\[-8pt]
\multicolumn{9}{l}{\footnotesize  $^*I_{a2}$ is the axial moment of inertia of the methyl group,}\\
\multicolumn{9}{l}{\footnotesize $I_{a1}$ is that of the hydroxyl group, $I_{a}$ = $I_{a1}$ + $I_{a2}$.}
\end{tabular}
\end{table}

\begin{table*}[htbp]
\caption{Spectroscopic parameters of the RF Hamiltonian 
and the asymmetry parameter $\kappa$ 
for CH$_3$OD, CD$_3$OH, CD$_3$OD, and CH$_3$OH.}
\label{T2}
\begin{tabular}{ l r@.l r@.l r@.l r@.l }
\hline\\[-5pt]
 & \multicolumn{2}{c}{CH$_3$OD} & \multicolumn{2}{c}{CD$_3$OH}
& \multicolumn{2}{c}{CD$_3$OD} & \multicolumn{2}{c}{CH$_3$OH}   \\
\hline\\[-7pt]
$A$ (cm$^{-1}$) & 3&679363 & 2&361733  & 2&169903& 4&255604\\
$B$ (cm$^{-1}$) & 0&7818695  & 0&6620463   & 0&6302237& 0&8232282\\
$C$ (cm$^{-1}$) & 0&7333172  & 0&6429177   & 0&5987836& 0&7928686\\
$D$ (cm$^{-1}$) & 0&02923331   &$-0$&003250156    &0&01897685&$-$0&002613155\\
$F$ (cm$^{-1}$) & 17&58161 & 25&09238&14&81704&27&75181\\
$V_3$ (cm$^{-1}$) & 370&3& 371&8 & 362&8& 375&6\\
$\rho$ & 0&7031751& 0&8948290 & 0&8224358& 0&8109008\\
$\kappa$ & $-0$&967 & $-0$&978 & $-0$&960& $-0$&982\\
\hline\\[-8pt]
\multicolumn{9}{l}{\footnotesize Values of $V_3$ are taken from [47].}
\end{tabular}
\end{table*}

The degree of asymmetry is defined by the parameter $\kappa$:
\begin{equation}
\kappa =  \frac{2B-A-C}{A-C}\ ,
\label{Eq9}
\end{equation}
which is equal to 1 or $-1$ when the molecule is symmetric ([48]).
Table~\ref{T2} contains the calculated spectroscopic parameters
for CH$_3$OD, CD$_3$OH and CD$_3$OD with the potential barriers $V_3$
and the asymmetry parameter $\kappa$.

The calculated spectroscopic parameters were used to determine the Hamiltonian matrix,
the non-vanishing elements of which are listed in [22]
for specific values of the rotational angular momentum $J$.
Diagonalization of this matrix gives simultaneously
the eigen-energies and eigenfunctions for $A$- and $E$-type methanol. 
The dependence of the found eigenvalues $E_i$ on $\Delta\mu/\mu$
defines the sensitivity coefficient $Q_{\mu}$ through the following relations:
\begin{equation}
\Delta E_i = q_i\frac{\Delta\mu}{\mu},
\label{Eq14}
\end{equation}
and
\begin{equation}
Q_\mu = \frac{q_{\scriptscriptstyle u} - q_{\scriptscriptstyle \ell}}{\nu}.
\label{Eq15}
\end{equation}
Here $q_i$ is the so-called $q$-factor, individual for each level,
which shows a response of the level $E_i$ to a small change of $\mu$, when
$|\Delta\mu/\mu| \ll 1$.
This functional dependence is assessed through the diagonalization of the 
Hamiltonian matrix for three sets of parameters that correspond to
$\mu = \mu_{\scriptscriptstyle 0}$ and 
$\mu = \mu_{\scriptscriptstyle 0}(1 \pm \varepsilon)$, where $\varepsilon$
is equal to 0.001 or 0.0001, and $\mu_{\scriptscriptstyle 0} \equiv 
\mu_{\scriptscriptstyle\rm lab}$ ([12]).
The subscripts $u$ and $\ell$ denote the upper and lower levels, respectively.

For calculations we selected molecular transitions with $0 \le J \le 13$
and $|\Delta K| = 1$,
since transitions with $\Delta K = 0$ are purely rotation
and have $Q_{\mu} \approx 1$.
Our calculations for the three deuterated isotopologues of methanol CH$_3$OD, CD$_3$OH and CD$_3$OD 
cover the low frequency range $1-50$~GHz which contains transitions with
the highest sensitivity to $\mu$-variations.

The results of our calculations are reported in
Table~\ref{T3} (CH$_3$OD)\footnote{The sensitivity coefficients
for CH$_3$OD are calculated for the first time.}, 
Table~\ref{T4} (CD$_3$OH), and Table~\ref{T5} (CD$_3$OD).
The structure of these tables is the following. 
The first column contains the quantum numbers $J$ and $K$ of the upper and lower levels. 
The second and third columns show the rest frequencies.
The fourth column lists the sensitivity coefficients $Q_\mu$ and their uncertainties,
estimated according to the scheme described in Section~\ref{Sec3}.
Included in these tables are 
molecular transitions with large sensitivity coefficients, $|Q_\mu| > 3$,
except for the $8_{-1}-8_0 E$ transition in CD$_3$OD with $Q_\mu = -0.9$ which is added to Table~\ref{T5}
for comparison with [13].

Tables~\ref{T3}--\ref{T5} show that deuterated isotopologues of methanol have 
high sensitivity coefficients of both signs.
The calculated values of $Q_\mu$ range from $-32$ to +25 for CH$_3$OD, 
from $-300$ to +73 for CD$_3$OH, and from
$-44$ to +38 for CD$_3$OD. 
The results obtained indicate that
the most attractive molecule for testing $\mu$-variations is CD$_3$OH, 
for which the maximum difference between the sensitivity coefficients 
(denominator in Eq.~\ref{Eq1-1})
reaches the value of
$\Delta Q_\mu \approx 370$ in the narrow frequency range $1-5$~GHz which 
by itself minimizes possible
systematic errors in the position measuring of spectral lines from a wide frequency range.

\begin{table*}[htbp]
\caption{Numerical calculation of the sensitivity coefficients $Q_\mu$ for 
the low-frequency torsion-rotation transitions ($\Delta K=\pm 1$) in CH$_3$OD.
Given in parenthesis are error estimates in the last digits.
}
\label{T3}
\begin{tabular}{l r@{.}l r@{.}l c |  l r@{.}l r@{.}l c }
\hline
\\[-10pt]
 \multicolumn{1}{c}{Transition} & \multicolumn{4}{c}{Frequency (MHz)}&\multicolumn{1}{c}{$Q_\mu$}
 &
 \multicolumn{1}{|c}{Transition} & \multicolumn{4}{c}{Frequency (MHz)}&\multicolumn{1}{c}{$Q_\mu$} \\
\multicolumn{1}{c}{$J_{{\scriptscriptstyle u}K_{\scriptscriptstyle u}} \to 
J_{{\scriptscriptstyle \ell}K_{\scriptscriptstyle \ell}}$} & 
\multicolumn{2}{c}{Exper.} & \multicolumn{2}{c}{Theor.} & &
\multicolumn{1}{|c}{$J_{{\scriptscriptstyle u}K_{\scriptscriptstyle u}} \to
J_{{\scriptscriptstyle \ell}K_{\scriptscriptstyle \ell}}$} 
&\multicolumn{2}{c}{Exper.} & \multicolumn{2}{c}{Theor.} \\[1pt]
\hline
$11_{4}-12_{3}\ A^{-}$ & 3983&678$^*$ & 4839&8 & $-32(5)$ &
$7_{-1}-7_{0}\ E$ & 17889&040& 17982&9 & $-6.4(3)$ \\
$3_{0}-2_{1}\ A^{+}$ & 4812&471$^*$ & 4468&3 & $-19.3(1.0)$ &
$6_{-1}-6_{0}\ E$ & 18454&760 & 18525&9 & $-6.5(3)$ \\
$10_{-3}-11_{-2}\ E$ & 6346&731$^*$ & 7012&6 & $25(2)$ &
$5_{-1}-5_{0}\ E$ & 18792&970 & 18844&2 & $-6.6(3)$ \\
$9_{2}-8_{3}\ A^{-}$ & 6362&584$^*$ & 7195&5 & $15(2)$ &
$4_{-1}-4_{0}\ E$ & 18957&170 & 18992&2 & $-6.6(3)$ \\
$9_{2}-8_{3}\ E$ & 6761&206$^*$ & 5619&9 & $11(2)$ &
$1_{-1}-1_{0}\ E$ & 18957&950 & 18965&0 & $-6.8(3)$ \\
$9_{2}-8_{3}\ A^{+}$ & 10279&780& 11079&8 & $9.3(1.4)$ &
$2_{-1}-2_{0}\ E$ & 18991&670 & 19004&8 & $-6.8(3)$ \\
$12_{-1}-12_{0}\ E$ & 11367&779$^*$ & 11580&5 & $-6.4(6)$ &
$3_{-1}-3_{0}\ E$ & 19005&640 & 19028&0 & $-6.7(3)$ \\
$11_{-1}-11_{0}\ E$ & 13033&065$^*$ & 13225&9 & $-6.4(5)$ &
$1_{1}-2_{0}\ E$ & 19518&790 & 19404&9 & $3.67(17)$ \\
$10_{-1}-10_{0}\ E$ & 14588&875$^*$ & 14758&9 & $-6.4(4)$ &
$6_{2}-7_{1}\ A^{+}$ & 23407&520& 23200&3 & $6.6(6)$ \\
$3_{2}-4_{1}\ E$ & 14920&430& 14716&3 & $-8.8(6)$ &
$5_1-4_{2}\ E $& 30839&200 & 31093&0 & $5.7(3)$\\
$7_{1}-6_{2}\ A^{-}$ & 15467&910& 15510&4 & $-7.5(8)$ &
$5_2-6_{1}\ A^{-} $& 34537&290 & 34559&4 & $4.8(3)$\\
$8_{1}-7_{2}\ A^{+}$ & 15720&990& 16043&7 & $-7.3(1.0)$ &
$1_1-2_{0}\ A^+ $& 41858&774$^*$  & 42261&2 & $3.34(11)$\\
$9_{-1}-9_{0}\ E$ & 15948&140 & 16094&5 & $-6.4(3)$ &
$9_{-3}-10_{-2}\ E$ & 51926&768$^*$ & 52658&7 & $3.9(3)$ \\
$8_{-1}-8_{0}\ E$ & 17057&090 & 17176&5 & $-6.4(3)$\\ 
\hline\\[-8pt]
\multicolumn{12}{l}{\footnotesize {\it Notes:}
Experimental frequencies are taken from [49];
predicted frequencies are marked by asterisk.}
\end{tabular}
\end{table*}

\begin{table*}[htbp]
\caption{
Numerical calculation of the sensitivity coefficients $Q_\mu$ for 
the low-frequency torsion-rotation transitions ($\Delta K=\pm 1$) in CD$_3$OH.
Given in parenthesis are error estimates in the last digits.
}
\label{T4}
\begin{tabular}{l r@{.}l r@{.}l c  |  l r@{.}l r@{.}l c }
\hline
\\[-10pt]
\multicolumn{1}{c}{Transition} & \multicolumn{4}{c}{Frequency (MHz)}&\multicolumn{1}{c}{$Q_\mu$}
&
\multicolumn{1}{|c}{Transition} & \multicolumn{4}{c}{Frequency (MHz)}&\multicolumn{1}{c}{$Q_\mu$}\\
\multicolumn{1}{c}{$J_{{\scriptscriptstyle u}K_{\scriptscriptstyle u}} \to 
J_{{\scriptscriptstyle \ell}K_{\scriptscriptstyle \ell}}$} &\multicolumn{2}{c}{Exper.} & \multicolumn{2}{c}{Theor.} &&
\multicolumn{1}{|c}{$J_{{\scriptscriptstyle u}K_{\scriptscriptstyle u}} \to
J_{{\scriptscriptstyle \ell}K_{\scriptscriptstyle \ell}}$} 
&\multicolumn{2}{c}{Exper.} & \multicolumn{2}{c}{Theor.} \\[1pt]
\hline
$2_{2}-1_{1}\ E$ & 1202&3089$^*$ & 1344&8 & $-300(12)^a$  &
$11_{3}-11_{2}\ A^{-+}$ & 17918&90 &18210&8 & $-20.2(9)$  \\
$1_{1}-2_{0}\ E$ & 1424&1693$^*$ & 1330&4 & $47(2)^b$  &
$9_{3}-9_{2}\ A^{+-}$ & 18081&20  &18289&3 & $-20.4(9)$  \\
$12_{-7}-13_{-6}\ E$ & 2656&1884$^*$ & 2032&2 & $-103(9)$ &
$12_{3}-12_{2}\ A^{-+}$ & 18097&00  &18436&6 & $-19.8(9)$  \\
$10_{-2}-9_{-3}\ E$ & 2754&4417$^*$& 3247&8 & $-84(5)$  &
$13_{3}-13_{2}\ A^{-+}$ & 18317&15  &18706&1 & $-19.4(9)$ \\
$8_{3}-9_{2}\ E$ & 2827&2446$^*$ & 2367&6 & $11(5)^c$ & 
$10_{3}-10_{2}\ A^{+-}$ & 18395&60  &18650&2 & $-19.9(9)$  \\
$6_{0}-5_{1}\ A^+$ & 2971&0289$^*$ & 3360&0 & $-96(4)^d$  &
$8_{6}-9_{5}\ A^{+}$ & 18988&36  &18429&6 & $-16.9(1.1)$ \\
$6_{1}-5_{2}\ A^-$ & 3130&6032$^*$  & 3417&2 & $-46(3)$ &
$8_{6}-9_{5}\ A^{-}$ & 18988&36  &18429&6 & $-16.9(1.1)$ \\
$3_{0}-4_{-1}\ E$ & 4011&3609$^*$ & 3734&5 & $73(3)^e$ &
$12_{3}-12_{2}\ A^{+-}$ & 19294&50  &19651&5 & $-18.7(8)$ \\
$11_{-5}-12_{-4}\ E$ & 4608&4386$^*$ & 4030&4 & $5(3)$ &
$13_{3}-13_{2}\ A^{+-}$ & 19905&8  &20318&1 & $-17.9(8)$ \\
$5_{5}-6_{4}\ E$ & 6723&8853$^*$ & 6293&0 & $-56(3)^f$  &
$10_{5}-9_{6}\ A^{+}$ & 20088&47& 20678&1 & $17.9(1.1)^g$ \\
$5_{2}-6_{1}\ A^+$ & 7662&4811$^*$ & 7379&8 & $20.2(1.3)$ &
$10_{5}-9_{6}\ A^{-}$ & 20088&47& 20678&1 & $17.9(1.1)^h$ \\
$8_{-1}-7_{-2}\ E$ & 9945&16& 10418&8 & $-22.0(1.1)$ &
$12_4-13_{3}\ E $& 24612&81  & 23996&1 & $13.5(7)$\\
$11_{3}-10_{4}\ A^+$ & 11125&37  &11672&4 & $-6.5(1.2)$  &
$9_{4}-10_{3}\ A^- $& 27981&73& 27515&2 & $4.0(5)$\\
$11_{3}-10_{4}\ A^-$ & 11215&14 &11764&1 & $-6.5(1.2)$ &
$9_{4}-10_{3}\ A^+ $& 28032&94& 27567&5 & $4.0(5)$\\
$3_{3}-3_{2}\ A^{-+}$ & 17328&19  &17340&1 & $-21.9(1.0)$  &
$6_{-2}-7_{-1}\ E $& 28811&25 & 28420&8 & $8.9(4)$\\
$3_{3}-3_{2}\ A^{+-}$ & 17335&14  &17347&0 & $-21.9(1.0)$ &
$7_{1}-6_{2}\ A^+ $& 29550&25 & 29899&6 & $-4.0(4)$\\
$4_{3}-4_{2}\ A^{-+}$ & 17353&94  &17387&0 & $-21.8(1.0)$ &
$7_{4}-6_{5}\ E $& 32352&307 & 32816&1 & $12.8(6)$\\
$4_{3}-4_{2}\ A^{+-}$ & 17374&42  &17407&7 & $-21.8(1.0)$ &
$5_{-1}-4_{0}\ E $& 35748&571 & 36099&7 & $-6.2(3)$\\
$5_{3}-5_{2}\ A^{-+}$ & 17388&63  &17447&6 & $-21.7(9)$ &
$8_{-3}-9_{-2}\ E $& 36663&008 & 36258&2 & $7.4(4)$\\
$6_{3}-6_{2}\ A^{-+}$ & 17433&99  &17523&4 & $-21.5(9)$ &
$4_{1}-5_{0}\ A^+ $& 37322&977 & 37013&0 & $8.7(3)$\\
$5_{3}-5_{2}\ A^{+-}$ & 17435&92  &17495&7 & $-21.6(9)$ &
$4_{2}-5_{1}\ A^- $& 37463&512& 37242&4 & $5.0(2)$\\
$7_{3}-7_{2}\ A^{-+}$ & 17492&27  &17616&0 & $-21.3(9)$ &
$3_{2}-2_{1}\ E $& 40649&426& 40858&1 & $-7.1(3)$\\
$6_{3}-6_{2}\ A^{+-}$ & 17527&61  &17618&3 & $-21.4(9)$ &
$11_{-7}-12_{-6}\ E $& 41737&585& 41149&7 & $-5.6(6)$\\
$8_{3}-8_{2}\ A^{-+}$ & 17566&15 &17727&7 & $-21.1(9)$ &
$2_{0}-3_{-1}\ E $& 43453&48 & 43245&6 & $8.5(3)$\\
$7_{3}-7_{2}\ A^{+-}$ & 17658&70  &17784&6 & $-21.1(9)$ &
$7_{0}-6_{1}\ A^+ $& 43501&125 & 43975&9 & $-5.6(3)$\\
$9_{3}-9_{2}\ A^{-+}$ & 17659&00  &17861&6 & $-20.8(9)$ &
$4_{2}-5_{1}\ A^+ $& 45160&765 & 44941&7 & $4.3(2)$\\
$10_{3}-10_{2}\ A^{-+}$ & 17774&89  &18021&1 & $-20.5(9)$ &
$9_{-1}-8_{-2}\ E $& 48567&661& 49130&7 & $-3.7(2)$\\
$8_{3}-8_{2}\ A^{+-}$ & 17839&40 &18004&7 & $-20.8(9)$ \\
\hline\\[-8pt]
\multicolumn{12}{l}{\footnotesize {\it Notes:}
Experimental frequencies are taken from [50];
predicted frequencies are marked by asterisk.}\\
\multicolumn{12}{l}{\footnotesize $Q_\mu$ from [21]:
$-330 \pm 20^a$,
$42 \pm 2^b$, $-43 \pm 2^c$, $-93 \pm 5^d$, $73 \pm 4^e$, $-167 \pm 8^f$, $88 \pm 4^g$, $88 \pm 4^h$.}\\
\end{tabular}
\end{table*}

\begin{table*}[htbp]
\caption{
Numerical calculation of the sensitivity coefficients $Q_\mu$ for 
the low-frequency torsion-rotation transitions ($\Delta K=\pm 1$) in CD$_3$OD.
Given in parenthesis are error estimates in the last digits.
}
\label{T5}
\begin{tabular}{l r@{.}l r@{.}l c  | l r@{.}l r@{.}l c }
\hline
\\[-10pt]
 \multicolumn{1}{c}{Transition} & \multicolumn{4}{c}{frequency (MHz)}&\multicolumn{1}{c}{$Q_\mu$}
&
\multicolumn{1}{|c}{Transition} & \multicolumn{4}{c}{Frequency (MHz)}&\multicolumn{1}{c}{$Q_\mu$}\\
\multicolumn{1}{c}{$J_{{\scriptscriptstyle u}K_{\scriptscriptstyle u}} \to 
J_{{\scriptscriptstyle \ell}K_{\scriptscriptstyle \ell}}$} &\multicolumn{2}{c}{Exper.} & \multicolumn{2}{c}{Theor.} &&
\multicolumn{1}{|c}{$J_{{\scriptscriptstyle u}K_{\scriptscriptstyle u}} \to
J_{{\scriptscriptstyle \ell}K_{\scriptscriptstyle \ell}}$} 
&\multicolumn{2}{c}{Exper.} & \multicolumn{2}{c}{Theor.} \\[1pt]
\hline
$1_{-1}-1_{0}\ E$ & 2237&8224$^*$ & 2229&4 & $-44.1(1.7)^a$  &
$4_{3}-5_{2}\ A^{-}$ & 5276&6275$^*$& 5020&4 & $-19.7(1.6)$ \\

$8_{-3}-7_{-4}\ E$ & 2328&9943$^*$ & 2699&6 & $38(5)^b$  &
$12_{0}-12_{-1}\ E$ & 5432&1218$^*$& 5456&3 & $9.3(3)$  \\

$2_{-1}-2_{0}\ E$ & 2476&7516$^*$ & 2472&2 & $-35.3(1.3)$ &
$13_{0}-13_{-1}\ E$ & 5898&7376$^*$  & 5913&7 & $9.5(3)$ \\

$3_{-1}-3_{0}\ E$ & 2777&7033$^*$ & 2778&4 & $-27.0(1.0)$  &
$1_{1}-2_{0}\ A^+$ & 8147&610 & 8072&4 & $11.5(4)$ \\

$4_{-1}-4_{0}\ E$ & 3100&6415$^*$  & 3107&6 & $-20.0(8)$ & 
$5_{1}-4_{2}\ A^+$ & 11755&04 & 11985&1 & $-4.4(6)$ \\

$5_{-1}-5_{0}\ E$ & 3413&586$^*$ & 3427&9 & $-14.4(6)$ &
$3_{2}-4_{1}\ A^-$ & 14172&74& 14028&4 & $5.5(4)$ \\

$9_{4}-10_{3}\ E$ & 3561&2687$^*$ & 3225&4 & $32(3)$  &
$3_{2}-4_{1}\ A^+$ & 22792&07 & 22615&8 & $3.8(3)$  \\

$6_{-1}-6_{0}\ E$ & 3695&4856$^*$ & 3717&7 & $-9.5(4)$ &
$11_{5}-12_{4}\ A^+$ & 24130&44 & 23768&0 & $5.7(6)$  \\

$7_{-1}-7_{0}\ E$ & 3940&3679$^*$ & 3970&3 & $-5.1(3)$ & 
$11_{5}-12_{4}\ A^-$ & 24141&87 & 23779&3 & $5.7(6)$  \\

$8_{-1}-8_{0}\ E$ & 4161&7206$^*$ & 4197&5 & $-0.9(2)^c$ & 
$6_{2}-5_{3}\ A^{-}$ & 31381&11& 31673&5 & $4.5(3)$ \\

$13_{5}-12_{6}\ E$ & 4254&1923$^*$& 4971&5 & $23(4)$ &
$6_{2}-5_{3}\ A^{+}$ & 31980&59& 32269&2 & $4.4(3)$ \\

$4_{3}-5_{2}\ A^{+}$ & 4977&3938$^*$& 4723&0 & $-20.9(1.7)$ &
$9_{-3}-8_{-4}\ E $& 39152&561 & 39556&0 & $3.2(3)$\\

$11_{0}-11_{-1}\ E$ & 5012&0901$^*$  &5044&5 & $8.3(3)$ &
$8_4-9_{3}\ E $& 40645&642  & 40372&7 & $3.8(3)$\\

$11_{4}-10_{5}\ E $& 5049&9782$^*$ & 5477&1 & $11(3)$\\
\hline\\[-8pt]
\multicolumn{12}{l}{\footnotesize {\it Notes:}
Experimental frequencies are taken from [41];
predicted transition frequencies are marked by asterisk.}\\
\multicolumn{12}{l}{\footnotesize $Q_\mu = -45 \pm 2^a, 80 \pm 4^b$ ([21]),
$Q_\mu = -0.7^c$ ([13]).
}\\
\end{tabular}
\end{table*}

\begin{table}
\begin{center}
\caption{The sensitivity coefficients for CH$_3$OH calculated with BELGI code, 
$Q_\mu^{\scriptscriptstyle\rm BELGI}$,
and ``toy'', $Q_\mu^{\scriptscriptstyle\rm toy}$, 
model from \citet{JKX11},
and, for comparison, 
the present paper $Q_\mu$ values.
Given in parentheses are errors in the last digits.}
\label{T6}
\begin{tabular}{l c c c c}
\hline
\\[-8pt]
Transition& Frequency$^*$ & $Q_\mu^{\scriptscriptstyle\rm BELGI}$
&$Q_\mu^{\scriptscriptstyle\rm toy}$&$Q_\mu$\\
{$J_{{\scriptscriptstyle u}K_{\scriptscriptstyle u}} \to 
J_{{\scriptscriptstyle \ell}K_{\scriptscriptstyle \ell}}$} 
&(MHz)&&\\
\hline
$5_{1} - 6_{0} A^+$&6668.52&42(2)&46&43(2)\\
$9_{-1} - 8_{-2} E$&9936.20&$-$11.5(6)&$-$16.7&$-$14.6(1.3)\\
$4_{3} - 5_{2} A^+$&9978.69&$-$53(3)&$-$35&$-$42(2)\\
$4_{3} - 5_{2} A^-$&10058.26&$-$52(3)&$-$35&$-$42(2)\\
$2_{0} - 3_{-1} E$&12178.59&33(2)&32&32.5(1.3)\\
$2_{1} - 3_{0} E$&19967.40&5.9(3)&5.0&6.3(3)\\
$9_{2} - 10_{1} A^+$&23121.02&11.7(6)&10.8&13.6(1.0)\\
$3_{2} - 3_{1} E$&24928.72&$-$17.9(9)&$-$15.2&$-$16.7(7)\\
$2_{2} - 2_{1} E$&24934.38&$-$17.9(9)&$-$15.2&$-$16.7(7)\\
$8_{2} - 9_{1} A^-$&28969.95&9.5(6)&8.8&11.0(7)\\
$4_{-1} - 3_{0} E$&36169.29&$-$9.7(5)&9.6&$-$9.6(4)\\
$6_{2} - 5_{3} A^+$&38293.29&15.1(8)&10.4&12.2(6)\\
$6_{2} - 5_{3} A^-$&38452.65&15.0(8)&10.4&12.2(6)\\
$7_{0} - 6_{1} A^+$&44069.48&$-$5.2(3)&$-$5.9&$-$5.4(3)\\
$1_{0} - 2_{-1} E$&60531.49&7.4(4)&7.3&7.3(3)\\
$1_{1} - 2_{0} E$&68305.68&2.4(1)&2.2&2.55(7)\\
\hline
\multicolumn{5}{l}{\footnotesize {\it Note:} $^*$Rest frequencies are taken from [51].}
\end{tabular}
\end{center}
\end{table}

\section{Estimation of errors of the sensitivity coefficients}
\label{Sec3}

Some of the sensitivity coefficients presented in Tables~\ref{T4} and \ref{T5}
were calculated earlier in [13], [21],
which makes it possible to compare the results of calculations using different forms
of the Hamiltonian.
The footnotes to these tables show that the $Q_\mu$ coefficients
from [13], [21]
differ significantly from our calculations for a number of lines. 
The most striking cases are marked by superscripts $c, f, g, h$ in Table~\ref{T4} and
by $b$ in Table~\ref{T5}. For the case $c$, we even found $Q_\mu$ of opposite signs.

\begin{figure}[htb]
\includegraphics[width=0.5\textwidth]{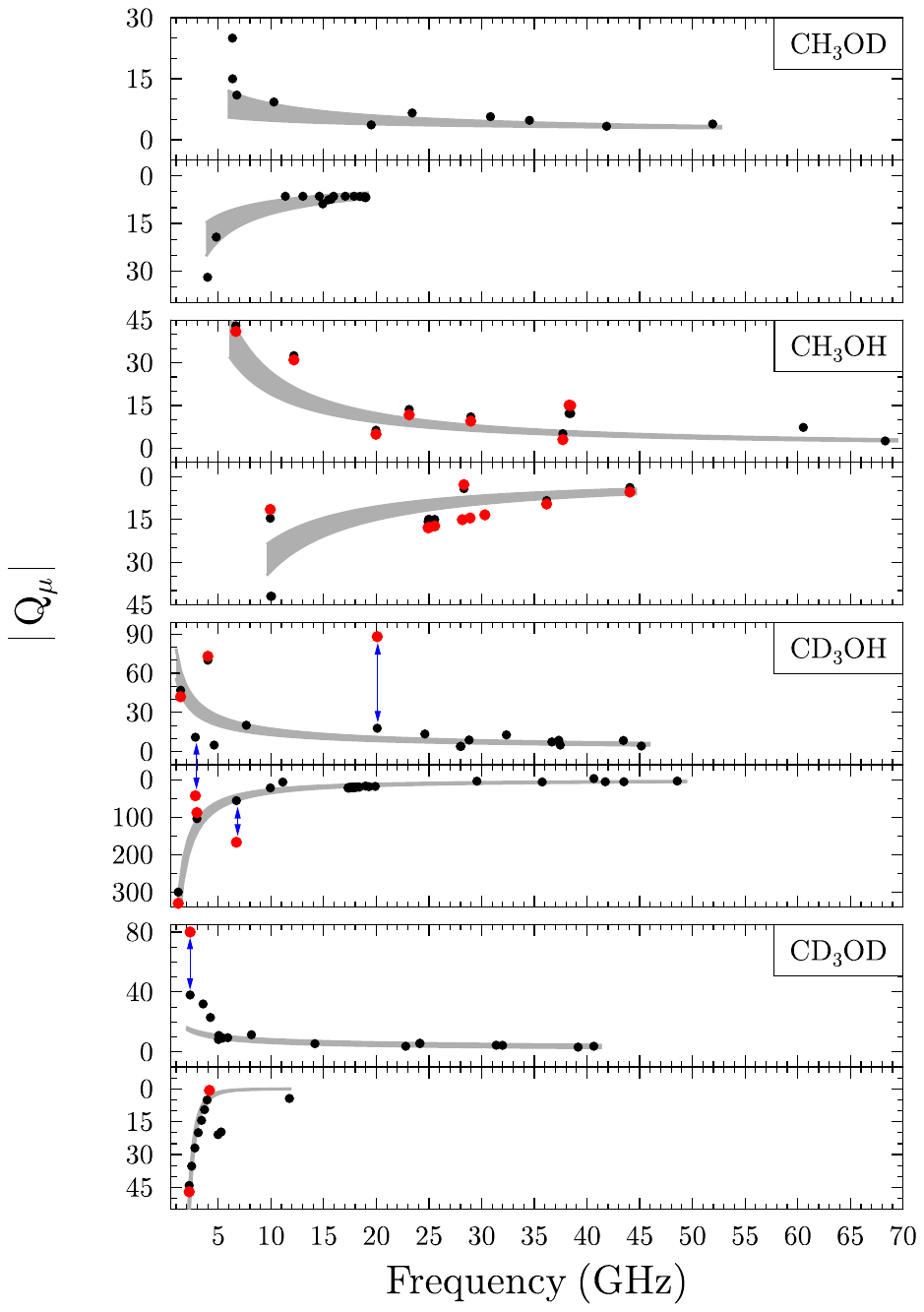}
\vspace{-2.0cm}
\caption{
Comparison of the sensitivity coefficients for 
methanol and its deuterated isotopologues: red dots~-- calculated by Jansen \etal\ [13], [21],
black dots~-- current results. The shadowed area shows the $\pm3\sigma$ uncertainty zone.
The blue arrows indicate maximum deviations in the $Q_\mu$ values between Jansen \etal\
and the present calculations. 
Note that $Q_\mu$ even changes sign in one case at 2.827~GHz for CD$_3$OH (see Table~\ref{T4}).
The isotopologue CH$_3$OD is considered for the first time in the present paper.
In general, $Q_\mu$ decreases with increasing frequencies of molecular transitions
in all four cases.
}
\label{F5}
\end{figure}

Figure~\ref{F5} illustrates the degree of agreement between common sensitivity coefficients
calculated in  [13], [21]
(red points) and in the present work (black points).
To understand large differences marked by the blue arrows in this figure,
we checked the errors of our procedure following the approach
developed in Refs.\ \ [12], [22]

Molecular transition frequencies depend on $\mu$ via the parameters
(\ref{Eq2})-(\ref{Eq8}) of the effective Hamiltonian.
Five of these parameters, $A$, \dots $F$, are inversely proportional to the moments of inertia,
which scales almost linearly with $\mu^{-1}$.
Thus, these parameters scale linearly with $\mu$.
The deviation from the linearity was estimated  
to be on the level 1-2\% ([12]).
It is mostly caused by the weak dependence of the internuclear distances on
$\mu$ due to the vibrational and centrifugal distortions.
Up to the similar small corrections, the remaining two parameters
($V_3$ and $\rho$) of the effective Hamiltonian 
do not depend on $\mu$. All these corrections change the scaling of the three terms in Eq.~(\ref{Eq2-1}). 
This, in turn, changes the sensitivity
coefficients in Eq.~(\ref{Eq2-5}). In order to estimate the scaling error for 
the sensitivity coefficients $Q_\mu$ we independently change the scaling of each of the parameters 
of the effective Hamiltonian by 2\% and calculate sensitivity coefficients using
Eqs.~(\ref{Eq14}, \ref{Eq15}).

Another possible source of theoretical errors comes from the interaction of the closely-lying 
torsion-rotation levels with the same exact quantum numbers. Such interaction leads to 
the well-known level repulsion.
Varying $\mu$ changes the distance between the levels and, thus, changes repulsion. 
This may affect the sensitivity coefficients of the interacting levels ([52]).
It is important that the $q$-factors change in the opposite directions leaving their sum unchanged.
Clearly, this source of errors is most important when there are close levels with 
significantly different $q$-factors.
Numerically it leads to the dependence of $q_i$, 
calculated from Eq.~(\ref{Eq15}), on $\mu$ ([53]).
To estimate this error, we calculated $q_i$ using $\Delta \mu=\pm \epsilon$ and estimated the
derivative $\partial q_i/\partial \mu$.
Our analysis shows that this source of errors is subdominant for all
levels considered in the present paper. 
The estimated theoretical errors are given in parenthesis in Tables~\ref{T3}-\ref{T6}.

Coming back to the comparison of different sets of the $Q_\mu$ values shown in Fig.~\ref{F5},
we note that in general there is a smooth dependence of $Q_\mu$ on frequency, despite some
scatter of points at low frequencies caused by interference of nearby energy levels.
The highest $Q_\mu$ values are found for transitions with the smallest differences
between the energy levels, $\Delta E \la 0.1$~K.
The results presented in Fig.~\ref{F5} show
a good concordance between [21]
calculations and ours for the parent
molecule CH$_3$OH, but essential deviations in some cases for its isotopologues.

Among them, the most striking difference occurs at 20.088~GHz ($10_5-9_6 A$)
in CD$_3$OH (Table~\ref{T4}) where Jansen's point lies too far from the $\pm3\sigma$ 
uncertainty zone shown by the shadowed area, 
taking into account that $\Delta E = 0.96$~K for this transition.

The red point at 6.723~GHz ($5_5-6_4 E$) also deviates from the general trend, while the black one does not. However, this deviation is comparable to the overall spread at $\Delta E \lesssim 0.3$~K.
As for the $Q_\mu$ values at 2.827~GHz ($8_3-9_2 E$), 
which have even opposite signs in our and Jansen's calculations,
we cannot state with certainty for the intermediate value of $\Delta E = 0.14$~K 
for this transition whose result is more reliable
since the found values of $Q_\mu$ can belong either to the upper or to the lower curve.
The isotopologue CD$_3$OD (Table~\ref{T5}) also demonstrates one point at 2.328~GHz 
($8_{-3}-7_{-4} E$) where our and Jansen's results differ significantly.

To illustrate the dependence of some values of the sensitivity coefficients on the
Hamiltonian model used, we compare in Table~\ref{T6} the 
computational results obtained by three models of the effective Hamiltonian
which are outlined below.

As mentioned above, [21]
used the complex model of the Hamiltonian
based on 119 parameters and calculated numerically the torsion-rotation levels with the BELGI code.
Despite the fact that this model calculates
the transition frequencies with high accuracy, it provides limited insight. 

For a clearer understanding of the physical meaning,
[13]
used a ``toy'' model with six parameters:
$A, B, C, F, \rho$ and $V_3$. 
Using this model, the authors calculated the sensitivity coefficients
for some transitions in CH$_3$OH. 
To compare with our results,
we recalculated the sensitivity coefficients $Q_\mu$ for the same transitions using 
the seven spectroscopic parameters from the last column in Table~\ref{T2}.
The result of such comparison is presented in Table~\ref{T6}.
As seen, there are some discrepancies between the listed $Q_\mu$ values,
the most pronounced of which are found for the following transitions:
$9_{-1}-8_{-2} E$, $4_{3}-5_{2} A^+$, $4_{3}-5_{2} A^-$,
$6_{2}-5_{3} A^+$, and $6_{2}-5_{3} A^-$.

The 119 parameters of the BELGI model have different dependence on $\mu$. 
Many of the higher-order parameters are products of operators and can be quite correlated.
Therefore, the exact relationship between the higher-order parameters of moments of inertia 
(and hence masses) is poorly known. 
It can give some effects in calculating the sensitivity coefficients
which were tested for CH$_3$OH in [13]
by using different forms of the effective
Hamiltonian. The authors showed that the calculated sensitivity coefficients coincide within
a few percent.
However, it is possible that the effects produced by the high-order parameters may provide a more significant
contribution for methanol deuterated isotopologues.
In any case, care should be taken with those $Q_\mu$ values where large deviations
have been detected when different forms of the effective Hamiltonian were used.
Elucidating the nature of the discrepancies found requires more detailed analysis.

\section{Conclusions}
\label{Sec4}

In this paper we calculated the sensitivity coefficients, $Q_\mu$, for different molecular
transitions from the microwave range 1--50~GHz to small changes in the values of 
the electron-to-proton mass ratio, $\mu = m_{\rm e}/m_{\rm p}$. 
The considered molecular species include CH$_3$OD, CD$_3$OH, and CD$_3$OD. 
The computational procedure is based on the effective Hamiltonian 
in the form suggested by [45].
The obtained main results are as follows:
\begin{enumerate}
\item
For CH$_3$OD, the $Q_\mu$ values lie in the interval from $-32$ to +25.

\item
For CD$_3$OH, the sensitivity coefficients vary in a wide range from $-300$ at 1.202~GHz
to +73 at 4.011~GHz.

\item
For CD$_3$OD, the two neighboring transitions at 2.237
($Q_\mu = -44$) and 2.328~GHz ($Q_\mu = +38$) confine the $Q_\mu$ values from the whole
sample in the frequency range 2--40~GHz.

\item
We also found that along with good agreement of the $Q_\mu$ values for the parent methanol CH$_3$OH
calculated with different Hamiltonians in the present and previous works, there are several pronounced outliers 
of unclear nature in the isotopologues CD$_3$OH and CD$_3$OD.
\end{enumerate}


\section*{Acknowledgements}
{J.S.V. and S.A.L. are supported in part by the Russian Science Foundation
under grant No.~23-22-00124.}

\section*{References}

[1] R. I. Thompson, The determination of the electron to
proton inertial mass ratio via molecular transitions, Astrophys. Lett. 16, 3 (1975).

[2] P. J. McMillan, Erratum: The mass distribution and
gravitational potential of the Milky Way, Monthly Notices of the Royal Astronomical Society 466, 174 (2017).

[3] J. Khoury and A. Weltman, Chameleon Fields: Awaiting
Surprises for Tests of Gravity in Space, Phys. Rev. Lett.
93, 171104 (2004), arXiv:astro-ph/0309300 [astro-ph].

[4] K. A. Olive and M. Pospelov, Environmental dependence of masses and coupling constants, Phys. Rev. D
77, 043524 (2008), arXiv:0709.3825.

[5] S. A. Levshakov, K. W. Ng, C. Henkel, and B. Mookerjea, [C I], [C II] and CO emission lines as a probe
for a variations at low and high redshifts, Monthly Notices of the Royal Astronomical Society 471, 2143 (2017),
arXiv:1707.03760 [astro-ph.GA].

[6] S. A. Levshakov, K. W. Ng, C. Henkel, B. Mookerjea, I. I.
Agafonova, S. Y. Liu, and W. H. Wang, Testing the weak
equivalence principle by differential measurements of fun-
damental constants in the Magellanic Clouds, Monthly
Notices of the Royal Astronomical Society 487, 5175
(2019), arXiv:1906.03682 [astro-ph.CO].

[7] S. A. Levshakov and D. A. Varshalovich, Molecular hy-
drogen in the z=2.811 absorbing material toward the
quasar PKS 0528-250, Monthly Notices of the Royal Astronomical Society 212, 517 (1985).

[8] D. A. Varshalovich and S. A. Levshakov, On a time dependence of physical constants, JETP Letters 58, 237
(1993).

[9] H. Rahmani, M. Wendt, R. Srianand, P. Noterdaeme,
P. Petitjean, P. Molaro, J. B. Whitmore, M. T. Murphy, M. Centurion, H. Fathivavsari, S. D'Odorico, T. M.
Evans, S. A. Levshakov, S. Lopez, C. J. A. P. Martins,
D. Reimers, and G. Vladilo, The UVES large program
for testing fundamental physics - II. Constraints on a
change in u towards quasar HE 0027-1836, Monthly Notices of the Royal Astronomical Society 435, 861 (2013),
arXiv:1307.5864 [astro-ph.CO].

[10] J. Bagdonaite, W. Ubachs, M. T. Murphy, and J. B.
Whitmore, Constraint on a Varying Proton-Electron
Mass Ratio 1.5 Billion Years after the Big Bang, Phys.
Rev. Lett. 114, 071301 (2015), arXiv:1501.05533 [astro-ph. CO].

[11] V. V. Flambaum and M. G. Kozlov, Limit on the cosmological variation of mp/me from the inversion spectrum of ammonia, Phys. Rev. Lett. 98, 240801 (2007),
arXiv:0704.2301.

[12] S. A. Levshakov, M. G. Kozlov, and D. Reimers,
Methanol as a tracer of fundamental constants, Astrophysical Journal 738, 26 (2011), arXiv:1106.1569.

[13] P. Jansen, I. Kleiner, L.-H. Xu, W. Ubachs, and H. L.
Bethlem, Sensitivity of Transitions in Internal Rotor
Molecules to a Possible Variation of the Proton-to-Electron Mass Ratio, Phys. Rev. A 84, 062505 (2011),
arXiv:1109.5076.

[14] N. Kanekar, W. Ubachs, K. M. Menten, J. Bagdonaite,
A. Brunthaler, C. Henkel, S. Muller, H. L. Bethlem, and
M. Dapra, Constraints on changes in the proton-electron
mass ratio using methanol lines, Monthly Notices of the
Royal Astronomical Society 448, L104 (2015), 1412.7757.

[15] M. Dapra, C. Henkel, S. A. Levshakov, K. M. Menten,
S. Muller, H. L. Bethlem, S. Leurini, A. V. Lapinov,
and W. Ubachs, Testing the variability of the proton-to-
electron mass ratio from observations of methanol in the
dark cloud core L1498, Monthly Notices of the Royal Astronomical Society 472, 4434 (2017), arXiv:1709.03103
[astro-ph.CO].

[16] S. Ellingsen, M. Voronkov, and S. Breen, Practical Limitations on Astrophysical Observations of Methanol to Investigate Variations in the Proton-to-Electron Mass Ratio, Phys. Rev. Lett. 107, 270801 (2011), arXiv:1111.4708
[astro-ph.CO].

[17] S. A. Levshakov, I. I. Agafonova, C. Henkel, K.-T.
Kim, M. G. Kozlov, B. Lankhaar, and W. Yang, Prob-
ing the electron-to-proton mass ratio gradient in the
Milky Way with Class I methanol masers, Monthly Notices of the Royal Astronomical Society 511, 413 (2022),
arXiv:2112.14560.

[18] S. A. Levshakov, D. Reimers, C. Henkel, B. Winkel,
A. Mignano, M. Centurión, and P. Molaro, Limits on
the spatial variations of the electron-to-proton mass ratio in the Galactic plane, Astronomy and Astrophysics
559, A91 (2013), arXiv:1310.1850 [astro-ph.GA].

[19] V. V. Ilyushin, P. Jansen, M. G. Kozlov, S. A. Levshakov,
I. Kleiner, W. Ubachs, and H. L. Bethlem, Sensitivity to
a possible variation of the proton-to-electron mass ratio
of torsion-wagging-rotation transitions in methylamine
CH3NH2, Phys. Rev. A 85, 032505 (2012).

[20] P. Jansen, L.-H. Xu, I. Kleiner, H. L. Bethlem, and
W. Ubachs, Methyl mercaptan (CH3SH) as a probe for
variation of the proton-to-electron mass ratio, Phys. Rev.
A 87, 052509 (2013), arXiv:1304.5249 [physics.atom-ph].

[21] P. Jansen, L.-H. Xu, I. Kleiner, W. Ubachs, and H. L.
Bethlem, Methanol as a sensitive probe for spatial and
temporal variations of the proton-to-electron mass ratio,
Phys. Rev. Lett. 106, 100801 (2011).

[22] J. S. Vorotyntseva, M. G. Kozlov, and S. A. Levshakov,
Methanol isotopologues as a probe for spatial and temporal variations of the electron-to-proton mass ratio,
Monthly Notices of the Royal Astronomical Society 527,
2750 (2024), arXiv:2310.04485.

[23] J.-H. Wu, X. Chen, Y.-K. Zhang, S. P. Ellingsen, A. M.
Sobolev, Z. Zhao, S.-M. Song, Z.-Q. Shen, B. Li, B. Xia,
R.-B. Zhao, J.-Q. Wang, and Y.-J. Wu, Physical En-
vironments of the Luminosity Outburst Source NGC
6334I Traced by Thermal and Maser Lines of Multiple
Molecules, Astrophysical Journal Supplement 265, 49
(2023).

[24] C. A. Gottlieb, J. A. Ball, E. W. Gottlieb, and D. F. Dickinson, Interstellar alcohol., Astrophysical Journal 227,
422 (1979).

[25] B. E. Turner, A Molecular Line Survey of Sagittarius B2
and Orion-KL from 70 to 115 GHz. II. Analysis of the
Data, Astrophysical Journal Supplement 76, 617 (1991).

[26] A. Belloche, H. S. P. Müller, R. T. Garrod, and K. M.
Menten, Exploring molecular complexity with ALMA
(EMoCA): Deuterated complex organic molecules in
Sagittarius B2(N2), Astronomy and Astrophysics 587,
A91 (2016), arXiv:1511.05721 [astro-ph.GA].

[27] R. Mauersberger, C. Henkel, T. Jacq, and C. M. Walmsley, Deuterated methanol in Orion., Astronomy and Astrophysics 194, L1 (1988).

[28] T. Jacq, C. M. Walmsley, R. Mauersberger, T. Anderson,
E. Herbst, and F. C. De Lucia, Detection of interstellar CH2 DOH., Astronomy and Astrophysics 271, 276
(1993).

[29] T. C. Peng, D. Despois, N. Brouillet, B. Parise,
and A. Baudry, Deuterated methanol in Orion
BN/KL, Astronomy and Astrophysics 543, A152 (2012),
arXiv:1206.2140 [astro-ph.GA].

[30] O. H. Wilkins and G. A. Blake, Relationship between
CH3OD Abundance and Temperature in the Orion KL
Nebula, Journal of Physical Chemistry A 126, 6473
(2022), arXiv:2208.08543 [astro-ph.GA].

[31] J. Ospina-Zamudio, C. Favre, M. Kounkel, L. H. Xu,
J. Neill, B. Lefloch, A. Faure, E. Bergin, D. Fedele, and
L. Hartmann, Deuterated methanol toward NGC 7538-
IRS1, Astronomy and Astrophysics 627, A80 (2019),
arXiv:1905.09798 [astro-ph.GA].

[32] E. G. Bogelund, B. A. McGuire, N. F. W. Ligterink,
V. Taquet, C. L. Brogan, T. R. Hunter, J. C. Pearson,
M. R. Hogerheijde, and E. F. van Dishoeck, Low levels of
methanol deuteration in the high-mass star-forming region NGC 6334I, Astronomy and Astrophysics 615, A88
(2018), arXiv:1804.01090 [astro-ph.GA].

[33] Y. C. Minh, Deuterated Methanol (CH3OD) in the
Hot Core of the Massive Star-Forming Region DR21
(OH), Publication of Korean Astronomical Society 29,
29 (2014).

[34] A. Ratajczak, V. Taquet, C. Kahane, C. Ceccarelli,
A. Faure, and E. Quirico, The puzzling deuteration of
methanol in low- to high-mass protostars, Astronomy and
Astrophysics 528, L13 (2011).

[35] B. Parise, C. Ceccarelli, A. G. G. M. Tielens, E. Herbst,
B. Lefloch, E. Caux, A. Castets, I. Mukhopadhyay, L. Pagani, and L. Loinard, Detection of doubly-deuterated
methanol in the solar-type protostar IRAS 16293-
2422, Astronomy and Astrophysics 393, L49 (2002),
arXiv:astro-ph/0207577 [astro-ph].

[36] J. K. Jorgensen, H. S. P. Müller, H. Calcutt, A. Coutens,
M. N. Drozdovskaya, K. I. Öberg, M. V. Persson,
V. Taquet, E. F. van Dishoeck, and S. F. Wampfler, The
ALMA-PILS survey: isotopic composition of oxygen-
containing complex organic molecules toward IRAS
16293-2422B, Astronomy and Astrophysics 620, A170
(2018), arXiv:1808.08753 [astro-ph.SR].

[37] B. Parise, C. Ceccarelli, A. G. G. M. Tielens, A. Castets,
E. Caux, B. Lefloch, and S. Maret, Testing grain sur-
face chemistry: a survey of deuterated formaldehyde and
methanol in low-mass class 0 protostars, Astronomy and
Astrophysics 453, 949 (2006), arXiv:astro-ph/0603135
[astro-ph].

[38] E. Bianchi, C. Codella, C. Ceccarelli, F. Fontani, L. Testi,
R. Bachiller, B. Lefloch, L. Podio, and V. Taquet, Decrease of the organic deuteration during the evolution of
Sun-like protostars: the case of SVS13-A, Monthly No-
tices of the Royal Astronomical Society 467, 3011 (2017),
arXiv:1701.08656 [astro-ph.SR].

[39] V. Taquet, E. Bianchi, C. Codella, M. V. Persson, C. Cec-
carelli, S. Cabrit, J. K. Jorgensen, C. Kahane, A. López-
Sepulcre, and R. Neri, Interferometric observations of
warm deuterated methanol in the inner regions of low-
mass protostars, Astronomy and Astrophysics 632, A19 (2019), arXiv:1909.08515 [astro-ph.GA].

[40] B. Parise, A. Castets, E. Herbst, E. Caux, C. Ceccarelli, I. Mukhopadhyay, and A. G. G. M. Tielens, First
detection of triply-deuterated methanol, Astronomy and
Astrophysics 416, 159 (2004), arXiv:astro-ph/0311038
[astro-ph].

[41] V. V. Ilyushin, H. S. P. Müller, J. K. Jorgensen,
S. Bauerecker, C. Maul, R. Porohovoi, E. A. Alekseev, O. Dorovskaya, F. Lewen, S. Schlemmer, and
R. M. Lees, Investigation of the rotational spectrum of
CD3OD and an astronomical search toward IRAS 16293-
2422, Astronomy and Astrophysics 677, A49 (2023),
arXiv:2307.07801 [astro-ph.SR].

[42] Y.-B. Duan and A. B. McCoy, Global Fit of Torsion-
Rotational Transitions in the First Three Torsional
States of CH 3OD, Journal of Molecular Spectroscopy
199, 302 (2000).

[43] B. Lankhaar, W. Vlemmings, G. Surcis, H. J. van
Langevelde, G. C. Groenenboom, and A. van der Avoird,
Characterization of methanol as a magnetic field tracer
in star-forming regions, Nature Astronomy 2, 145 (2018),
arXiv:1802.05764 [astro-ph.GA].

[44] S. V. Kalenskii and S. Kurtz, Analytical methods
for measuring the parameters of interstellar gas using methanol observations, Astronomy Reports 60, 702
(2016), arXiv:1710.07605 [astro-ph.GA].

[45] D. Rabli and D. R. Flower, The rotational structure of
methanol and its excitation
by helium, Monthly Notices
of the Royal Astronomical Society 403, 2033 (2010).

[46] R. M. Lees and J. G. Baker, Torsion-Vibration-Rotation
Interactions in Methanol. I. Millimeter Wave Spectrum,
J. Chemical Physics
48, 5299 (1968).

[47] R. M. Lees, Torsion-Vibration-Rotation Interactions in
Methanol. II. Microwave Spectrum of CD3OD, J. Chem-
ical Physics 56, 5887 (1972).

[48] C. Townes and A. Schawlow, Microwave Spectroscopy
(McGraw-Hill, New York, 1955).

[49] T. Anderson, R. L. Crownover, E. Herbst, and F. C. De
Lucia, The Laboratory Millimeter- and Submillimeter-
Wave Spectrum of CH 3OD, Astrophysical Journal Sup-
plement 67, 135 (1988).

[50] V. V. Ilyushin, H. S. P. Müller, J. K. Jorgensen,
S. Bauerecker, C. Maul, Y. Bakhmat, E. A. Alekseev,
O. Dorovskaya, S. Vlasenko, F. Lewen, S. Schlemmer,
K. Berezkin, and R. M. Lees, Rotational and rovibrational spectroscopy of CD3OH with an account of
CD3OH toward IRAS 16293-2422, Astronomy and Astrophysics 658, A127 (2022), arXiv:2111.09055 [astro-
ph. GA].

[51] F. J. Lovas, NIST Recommended Rest Frequencies
for Observed Interstellar Molecular Microwave Transitions 2002 Revision, Journal of Physical and Chemical
Reference Data 33, 177 (2004).

[52] V. A. Dzuba, V. V. Flambaum, M. G. Kozlov, and
M. Marchenko, a dependence of transition frequencies
for ions Si II, Cr III, Fe II, Ni II and Zn II, Phys. Rev. A
66, 022501 (2002), arXiv:physics/0112093.

[53] E. A. Konovalova, M. G. Kozlov, and R. T. Imanbaeva,
Coefficients of sensitivity to a-variation for astrophysically relevant transitions in ni ii, Phys. Rev. A 90, 042512
(2014), arXiv:1407.1860.

%

\end{document}